\documentstyle[12pt,psfig]{article}
\newcommand{\gtrsim}{ \mathop{}_{\textstyle \sim}^{\textstyle >} }
\newcommand{\lesssim}{ \mathop{}_{\textstyle \sim}^{\textstyle <} }
\newcommand{\eV}{~\mbox{eV}}

\newcommand{\GeV}{~\mbox{GeV}}
\newcommand{\TeV}{~\mbox{TeV}}

\newcommand{\gsim}{ \mathop{}_{\textstyle \sim}^{\textstyle >} }
\newcommand{\lsim}{ \mathop{}_{\textstyle \sim}^{\textstyle <} }
\newcommand{\vev}[1]{ \left\langle {#1} \right\rangle }
\newcommand{\dphi}{ \dot{\phi} }
\newcommand{\ddphi}{ \ddot{\phi} }
\newcommand{\Hoscth}{ H_{\rm osc}^{\rm th} }
\newcommand{\toscth}{ t_{\rm osc}^{\rm th} }
\def\Frac#1#2{{\displaystyle\frac{#1}{#2}}}
\begin{document}
\baselineskip 0.6cm
\renewcommand{\thefootnote}{\fnsymbol{footnote}}
\def\tr{\mathop{\rm tr}\nolimits}
\def\Tr{\mathop{\rm Tr}\nolimits}
\def\Re{\mathop{\rm Re}\nolimits}
\def\Im{\mathop{\rm Im}\nolimits}
\setcounter{footnote}{1}

\begin{titlepage}

\begin{flushright}
UT-904\\
\end{flushright}
 
\vskip 2cm
\begin{center}
 {\large \bf Affleck-Dine Leptogenesis with an Ultralight Neutrino}
 \vskip 1.2cm
 T. Asaka$^1$, Masaaki Fujii$^1$, K. Hamaguchi$^1$, and T. Yanagida$^{1,2}$

 \vskip 0.4cm

 {\it $^1$ Department of Physics, University of Tokyo,
 Tokyo 113-0033, Japan}\\
 {\it $^2$ Research Center for the Early Universe, University of Tokyo,
 Tokyo 113-0033, Japan}
\vskip 0.2cm
(August 4, 2000)
\vskip 2cm
\abstract{ We perform a detailed analysis on Affleck-Dine leptogenesis
taking into account the thermal effects on the dynamics of the flat
direction field $\phi$.  We find that an extremely small mass for the
lightest neutrino $\nu_1$, ${m_\nu}_1\lsim 10^{-8}$ eV, is required to
produce enough lepton-number asymmetry to explain the baryon asymmetry
in the present universe. We impose here the reheating temperature after
inflation $T_R$ to be $T_R\lsim 10^8$ GeV to solve the cosmological
gravitino problem. The required value of neutrino mass seems to be very
unlikely the case since the recent Superkamiokande experiments suggest the
masses of heavier two neutrinos $\nu_2$ and $\nu_3$ to be in a range of
$10^{-1}$--$10^{-3}$ eV.  We also propose a model to avoid this
difficulty based on the Peccei-Quinn symmetry, where the required
neutrino mass can be as large as ${m_\nu}_1\simeq 10^{-4}$ eV.  }
\end{center}
\end{titlepage}

\renewcommand{\thefootnote}{\arabic{footnote}}
\setcounter{footnote}{0}

%
\section{Introduction}
%
%
There have been growing interest in leptogenesis~\cite{1-Fuku-Yana} to
account for the baryon asymmetry in the present universe since the
Superkamiokande collaboration presented convincing evidence of the
atmospheric neutrino oscillation~\cite{2-Kamioka-atm}.  In fact, various
mechanisms~\cite{3-LG-Th,4-LG-ID,5-LG-AD} for the leptogenesis have been
proposed so far.  Among them the leptogenesis by flat directions
$\phi$ in supersymmetric (SUSY) theories, i.e., Affleck-Dine (AD)
mechanism~\cite{6-Affleck-Dine}, 
is expected to work even with relatively lower
reheating temperatures $T_R$ (e.g., $T_R \lsim 10^8$ GeV) and hence it
seems cosmologically safer than others, since the number density of
gravitinos produced in the reheating processes is proportional to the
reheating temperature $T_R$ of inflation~\cite{7-grav-prob}.
\footnote{ The gravitinos may be also produced
nonthermally~\cite{8-nonth-grav}.  We neglect such an effect in this
paper, since it depends on the details of inflation models.  }
However, it has been pointed
out~\cite{9-Dine-Rand-Thom,10-Alla-Camp-Elli} that the flat direction
fields $\phi$ start their coherent oscillations at an earlier time than
usually estimated due to interactions with thermal plasma of
inflaton-decay products and that the resultant baryon asymmetry is
strongly suppressed.

In this paper we perform a detailed numerical analysis on the AD
leptogenesis taking into account the thermal plasma effects.  It has
been argued in Ref.~\cite{9-Dine-Rand-Thom} that a relatively low
reheating temperature $T_R\lsim 10^6$ GeV is necessary to escape from
the thermal effects. We confirm that the baryon asymmetry is indeed
suppressed for $T_R\gsim 10^5$--$10^6$ GeV due to the thermal effects,
when the lightest neutrino mass is ${m_\nu}_1 \gtrsim 10^{-9}$ eV.  On
the other hand, we find that the thermal effects become significant for
a higher reheating temperature, $T_R \gtrsim 10^5$--$10^6$ GeV $\times
({m_\nu}_1 / 10^{-9} \eV )^{-1}$, if ${m_\nu}_1 \lesssim 10^{-9}$ eV.

Furthermore, we estimate how much the resultant baryon asymmetry is
suppressed for such a high reheating temperature by analytical and
numerical calculations.  What we have found is that an ultralight
neutrino of mass ${m_\nu}_1 \lsim 10^{-8}$ eV is required to obtain
the desired baryon asymmetry in the present universe, $n_B / s \simeq
(0.1$--$1) \times 10^{-10}$, when we impose the reheating temperature
$T_R$ to be $T_R\lsim 10^8$ GeV to avoid the the overproduction of
gravitinos~\cite{7-grav-prob}.  Here, $n_B$ and $s$ are the present
baryon-number and entropy densities, respectively.

We note, however, that the above neutrino mass is not necessarily the
same as the mass to be observed today, if effective Majorana masses for
right-handed neutrinos in the early universe are different from the
values in the true vacuum.  To demonstrate this point we show an
explicit model based on the Peccei-Quinn
symmetry~\cite{11-Peccei-Quinn}, which is one of the most attractive
solutions to the strong CP problem.  In this model we find that the mass
for the lightest neutrino in the true vacuum can be as large as
$10^{-4}$ eV. This is very close to the masses of heavier two neutrinos,
${m_\nu}_{2,3}\simeq 10^{-1}$--$10^{-3}$ eV, suggested from the recent
Superkaimokande experiments~\cite{2-Kamioka-atm,12-Suzuki}.

%
\section{Flat directions for the Leptogenesis}
\label{sec_FD}
%
%
We adopt flat directions $H_u = \widetilde{L_i} = \phi^i / \sqrt{2}$ (a
family index $i =$ 1, 2, 3) originally proposed by Murayama and one of
the authors (T.Y.)~\cite{13-Mura-Yana} as the Affleck-Dine flat direction
field for leptogenesis, where $\widetilde{L_i}$ are scalar components of
the chiral multiplets $L_i$ of $SU(2)_L$-doublet leptons.  We have
effective dimension-five operators in the superpotential,
\begin{eqnarray}
 W = \frac{1}{2 M_i}
  \left( L_i H_u \right)^2
  =
  \frac{{m_\nu}_i}{2 v_u^2}
  \left( L_i H_u \right)^2.
  \label{eq_super}
\end{eqnarray}
Here, $v_u \equiv \vev{H_u} = \sin\beta \times 174$ GeV and we have used
the seesaw formula~\cite{14-seesaw},
\footnote{$\tan \beta \equiv
\vev{H_u}/\vev{H_d}$, where $H_u$ and $H_d$ are Higgs supermultiplets
which couple to ``up''-type and ``down''-type quarks, respectively.  }
\begin{eqnarray}
 \label{eq_seesaw}
  {m_\nu}_i = \frac{\vev{H_u}^2}{M_i}.
\end{eqnarray}
Hereafter, we take $\sin\beta \simeq 1$, and we suppress the family
index $i$ for simplicity.  As we will see later, the relevant flat
direction $\phi$ for the most effective leptogenesis corresponds to the
first family (i.e., $\phi/\sqrt{2} \equiv \widetilde{L_1} = H_u$).  The
potential of the flat direction $\phi$ is given by
\begin{eqnarray}
 V = m^2|\phi|^2 
  + \frac{1}{8M}
  \left( A \phi^4 + A^* \phi^{*4} \right)
  + \frac{1}{4 M^2}
  |\phi|^6,
\end{eqnarray}
where $m$ and $A$ are SUSY-breaking mass parameters 
and the last term is derived directly from the
superpotential in Eq.~(\ref{eq_super}).
We choose $m\simeq |A| \simeq 1$ TeV.  

During inflation the energy density of the universe is dominated by the
inflaton $\chi$.  After the end of inflation the inflaton $\chi$ starts
its coherent oscillation and the energy density of the universe is
dominated by the inflaton $\chi$ also in this period.  
The non-zero vacuum energy of $\chi$ induces additional SUSY breaking effects.
Therefore, the flat direction $\phi$ receives an additional 
SUSY breaking mass of the order of $H$ in the $\chi$-dominated era.
Here $H$ is the Hubble parameter of the expanding universe.  There
is also an $A$-term proportional to the Hubble parameter, $V \sim ( H /
M ) \phi^4$.  Thus, we have the additional potential is given by
\begin{eqnarray}
 \delta V = - c_H H^2|\phi|^2
  + \frac{H}{8M}
  \left( a_H \phi^4 + a_H^* \phi^{*4} \right),
  \label{eq_potential_add}
\end{eqnarray}
where $c_H$ and $a_H$ are real and complex constants respectively, and we
take $c_H > 0$. (See the next section.)  Hereafter, we assume $c_H
\simeq |a_H| \simeq 1$.
\footnote{ The SUSY breaking effects due to the finite potential energy
of the inflaton $\chi$ depend on the details of the K\"ahler potential.
In fact, the coefficients of the Hubble-induced terms, $c_H$ and $a_H$
in Eqs.~(\ref{eq_potential_add}) and (\ref{eq_potential}), can be much
smaller than order unity.  However, we simply assume here $c_H \simeq
|a_H| \simeq 1$ since the conclusion does not depend much on these
parameters unless $c_H \ll -1$. (See, for example,
Ref.~\cite{15-Moro-Mura}.)  }

The crucial point in the following discussion is that the decay of the
inflaton $\chi$ occurs during the coherent oscillation of $\chi$, while
it completes much later than the beginning of the oscillation.  Thus,
even in the $\chi$ oscillation period, there is a dilute plasma with a
temperature $T = (T_R^2 M_* H)^{1/4}$ ($M_* = 2.4\times 10^{18}$ GeV
is the reduced Planck mass) arising from the inflaton
decay~\cite{16-Kolb-Turner}, although most of the energy density of the
universe is carried by the coherent oscillation.  Notice that in this
dilute plasma the temperature decreases as $T\propto H^{1/4}$ rather
than as $T\propto H^{1/2}$ in the usual radiation dominated universe.

It is quite plausible that the fields $\psi_k$ which couple to the
$\phi$ are produced by the inflaton decay and/or by thermal scatterings
if their effective masses $f_k |\phi|$ are smaller than the temperature
$T$,
\begin{eqnarray}
 f_k|\phi| < T,
  \label{eq_inTB}
\end{eqnarray}
and hence they must be included in the plasma.  Here, $f_k$ correspond 
to their Yukawa or gauge coupling constants of the flat direction $\phi$
(we take $f_k$ real and positive).

As stressed in Refs.~\cite{9-Dine-Rand-Thom,10-Alla-Camp-Elli} the
thermal effects from the dilute plasma may affect the dynamics of the
$\phi$ field.  In fact, the flat direction $\phi$ receives a thermal
mass of order $\sim f_k T$ if the condition Eq.~(\ref{eq_inTB}) is
satisfied.  Then the total effective potential for $\phi$ including the
Hubble-induced terms and the thermal mass term is given
by~\cite{10-Alla-Camp-Elli}
\begin{eqnarray}
 V_{\rm total} 
  &=& \left( m^2 - H^2 + \sum_{f_k|\phi| < T} c_k f_k^2 T^2 \right)
  |\phi|^2 
  \nonumber \\
 && + 
  \frac{m}{8M}
  \left(
   a \phi^4 + h.c.
   \right)
   + \frac{H}{8M}
   \left(
    a_H \phi^4 + h.c.
    \right)
    + \frac{1}{4 M^2}
  |\phi|^6,
  \label{eq_potential}
\end{eqnarray}
where $c_k$ are real positive constants of order unity and $a = A / m$
is a complex constant of order unity.  The summation of $f_k$ means that
only the fields in thermal plasma [i.e., their couplings $f_k$ satisfy
the condition Eq.~(\ref{eq_inTB})] can induce the thermal mass for
$\phi$.

We include all the couplings relevant for the flat direction $H_u =
\widetilde{L}_1 = \phi / \sqrt{2}$ in the SUSY standard model, i.e., the
gauge couplings for the $SU(2)_L$ and $U(1)_Y$ gauge groups, and Yukawa
couplings for up-type quarks and charged leptons.  The couplings $f_k$,
which are redefined so that the masses for the fields $\psi_k$ couple to
$\phi$ become $f_k|\phi|$, are listed in Table.\ref{table1} with the
coefficients $c_k$ of the thermal mass for $\phi$.
%
%
\begin{table}[t]
 \begin{center}
  \begin{tabular}{|c||c|c|c|c|}
   \hline $f_k$ (coupling for $\phi$) & $\sqrt{\Frac{g_1^2 + g_2^2}{2}}$
   & $\Frac{g_2}{\sqrt{2}}$ & $\Frac{y_a}{\sqrt{2}}$ ( $a = t, c, u$ )&
   $\Frac{y_{\l_1}}{\sqrt{2}}$ \\ \hline $c_k$ & $\Frac{1}{4}$ &
   $\Frac{1}{2}$ & $\Frac{3}{4}$ & $\Frac{1}{4}$ \\ \hline
  \end{tabular}
 \end{center}
 \caption{The couplings of the flat direction field $\phi$ to
 other fields and the coefficients $c_k$ of the thermal mass of $\phi$
 induced by these fields.
 $g_1$ and $g_2$ denote the gauge couplings for the $U(1)_Y$ and
 $SU(2)_L$ gauge group respectively, and $y_a$ ($a = t, c, u$) and
 ${y_l}_1$ are the Yukawa couplings for up-type quarks and charged lepton,
 respectively.}
 \label{table1}
\end{table}
%
%
Among the gauge supermultiplets, the flat direction $\phi$ does not
couple to one $U(1)$ group which remains unbroken by the condensate of
$\phi$.  Thus, there are one $Z$-like and two $W$-like massive gauge
multiplets as in the SUSY standard model.  For the couplings of the
charged leptons, we should be careful.  The flat direction $\phi$
receives its thermal mass from only one linear combination of charged
leptons.  The effective Yukawa coupling $y_{\l_1}$ is determined by the
Yukawa couplings for the charged leptons and the mixing matrix among
gauge eigenstates of the neutrinos.  (Note that the relevant $\phi$ is
the flat direction which corresponds to the first family $\nu_1$ of mass
eigenstates.)  Taking large mixing angles both for atmospheric and solar
neutrino oscillations indicated by the recent Superkamiokande
data~\cite{2-Kamioka-atm, 12-Suzuki}, we find ${y_l}_1 \simeq {\cal
O}(0.1$--$1)\times y_{\tau}$.  
Thus, $f_k$ are roughly $10^{-5}$--1 in the SUSY standard model.
We will see in Sec.~\ref{sec_lepton} that
the Yukawa coupling for the up quark induces relevant thermal effects,
while other couplings turn out to be too large to satisfy the condition
Eq.~(\ref{eq_inTB}) in most of the cases.

%
\section{Evolution of the flat direction}
\label{sec_evolution}
%
%
In this section we discuss the evolution of the flat direction
$\phi$, which is described by the equation of motion with the potential
$V_{\rm total}$ in Eq.~(\ref{eq_potential}),
\begin{eqnarray}
 \ddphi + 3 H \dphi + \frac{\partial V_{\rm total}}{\partial\phi^*} = 0,
  \label{eq_ofmotion}
\end{eqnarray}
where the dot denotes the derivative with time.

During inflation the Hubble parameter $H_{\rm inf}$ is much larger than
the soft SUSY breaking mass, $H_{\rm inf}\gg m$.  We assume that the
additional mass squared is negative ($c_H > 0$) and causes an
instability of $\phi$ at the origin, that is, $m_{\rm eff}^2 = m^2 - c_H
(H_{\rm inf})^2 < 0$. Then, the minima of the potential are given by
\begin{eqnarray} 
 |\phi| &\simeq& \sqrt{M H_{\rm inf}},
  \label{eq_minimum_inf1}
  \\
 \arg(\phi) &\simeq& \frac{- \arg(a_H) + ( 2 n + 1 )\pi}{4}
  \quad n = 0\cdots 3.
  \label{eq_minimum_inf2}
\end{eqnarray}
Note that there is no plasma and hence no thermal mass term during the
inflation.  Because curvatures of the potential around the minimum along
both the radius and phase directions are of the order of the Hubble
parameter $H_{\rm inf}$, the flat direction $\phi$ runs to one of the
four minima ($n = n_0$) from any given initial value and 
is settled down it.

After the inflation ends, the inflaton $\chi$ starts to
oscillate, and there appears a dilute plasma with a temperature $T =
(T_R^2 M_* H)^{1/4}$.  Then, the dynamics of the flat direction $\phi$
is determined by the total potential $V_{\rm total}$ in
Eq.~(\ref{eq_potential}).

In the period of the inflaton oscillation, the minimum of the $\phi$
potential moves toward the origin $\phi = 0$ as the Hubble parameter $H$
decreases.  As long as the potential for $\phi$ is dominated by
Hubble-induced (mass and $A$) terms and also $|\phi|^6$ term, the flat
direction field $\phi$ always tracks the following instantaneous minimum
of the potential~\cite{9-Dine-Rand-Thom,15-Moro-Mura},
\begin{eqnarray} 
 |\phi| &\simeq& \sqrt{MH},
  \\
 \arg(\phi) &\simeq& \frac{- \arg(a_H) + ( 2 n_0 + 1 )\pi}{4}.
\end{eqnarray}
This is because the curvatures around the minimum along
both radius and phase directions are of the order of $H$ also in this period,
and hence the $\phi$ always catches up the instantaneous
minimum.

As the universe evolves, the field value $|\phi|$ decreases with time.
When the field value $|\phi|$ satisfies the condition
Eq.~(\ref{eq_inTB}), the fields $\psi_k$ which couple to the $\phi$ can
enter in the dilute plasma so that they induce a thermal mass $c_k
f_k^2 T^2$ for $\phi$.  If this thermal mass exceeds the Hubble-induced
mass, i.e.,
\begin{eqnarray}
 H^2 < \sum_{f_k|\phi| < T} c_k f_k^2 T^2,
  \label{eq_Tdominate}
\end{eqnarray}
the dynamics of the $\phi$ fields is drastically changed.  The equation
of motion for $\phi$ at this epoch is approximately given by
\begin{eqnarray}
 \ddphi + 3 H \dphi + c_k f^2_k T^2 \phi  = 0.
\end{eqnarray}
This equation can be solved analytically and the solution is given by
\begin{eqnarray}
 \phi(t)
  &=&
  \phi_1
  \left[ \frac{1}{z^{2/3}} J_{2/3}(z) \right]
  +
  \phi_2
  \left[ \frac{1}{z^{2/3}} J_{-2/3}(z) \right]
  \nonumber
  \\
 z &=& \frac{4}{3}
  \left(
   \frac{2}{3} c_k^2 f^4_k M_* T_R^2 \,\,t^3
   \right)^{1/4},
   \label{eq_osc_phi}
\end{eqnarray}
where $\phi_1$ and $\phi_2$ are constants and $J_{\nu}$ is the Bessel
function.  We see that the $\phi$ oscillates around the origin $\phi
= 0$.  The time scale of the oscillation is $\sim \left(f^4_k M_* T_R^2
\right)^{-1/3}$ and the amplitude of the oscillation is damped as $|\phi|
\sim t^{-7/8} \sim H^{7/8}$.

Now let us estimate the cosmic time $\toscth$ when the flat direction
$\phi$ starts its oscillation due to the thermal mass term, since the
time $\toscth$ plays a crucial role in the production of the 
lepton-number asymmetry.  First of all, we
consider a simple case in which there is only one field $\psi_1$ with a
coupling $f_1$.  It is easily found from Eqs.~(\ref{eq_inTB}) and
(\ref{eq_Tdominate}) that the following two conditions must be satisfied
in order to cause the $\phi$'s oscillation,
\begin{eqnarray}
 H &<& \frac{1}{f_1^4}\frac{M_* T_R^2}{M^2},
  \label{eq_condition_1}
  \\
 H &<&  \left(c_1^2 f_1^4 M_* T_R^2 \right)^{1/3},
  \label{eq_condition_2}
\end{eqnarray}
where we have used the relations $|\phi|\simeq \sqrt{MH}$ and $T =
(T_R^2 M_* H)^{1/4}$.  If the coupling $f_1$ is small enough, the
$\psi_1$ can easily enter in the thermal plasma, but the induced thermal
mass for $\phi$ is smaller than the Hubble-induced mass $H$ at the
beginning.  However, since the temperature decreases more slowly than
the Hubble parameter $H$ ($T\propto H^{1/4}$) as the universe expands,
the thermal mass eventually exceeds $H$.  Then the $\phi$'s oscillation
starts at the time when the condition Eq.~(\ref{eq_condition_2}) is
satisfied.  On the other hand, if $f_1$ is very large, the would-be
thermal mass can be large enough to exceed the Hubble mass.  However,
the thermal mass does not appear until the $|\phi|$ becomes small enough
to satisfy Eq.~(\ref{eq_inTB}).  In this case the $\phi$ starts to
oscillate at the time when the condition Eq.~(\ref{eq_condition_1}) is
satisfied.  Therefore, we find the Hubble parameter $\Hoscth$ at $t =
\toscth$ to be
\begin{eqnarray}
 \Hoscth &=& {\rm min}\
  \left[ \frac{1}{f_1^4}\frac{M_* T_R^2}{M^2}\,\,,\,\,
   \left(c_1^2 f_1^4
    M_* T_R^2 \right)^{1/3}\right].
    \label{eq_Hosc_1}
\end{eqnarray}
Notice that in order to have a significant thermal effect, as discussed in
Ref.~\cite{10-Alla-Camp-Elli}, a coupling $f_1$ with intermediate
strength is necessary.  This is because too large $f_1$ can
not satisfy the condition Eq.~(\ref{eq_inTB}), while too small $f_1$ can
not give a large thermal mass for $\phi$ which satisfies the condition
Eq.~(\ref{eq_Tdominate}).

It is easy to apply the above discussion to the case of more than one
couplings.  If there is another coupling $f_i$ which can satisfy the
both conditions Eqs.~(\ref{eq_condition_1}) and (\ref{eq_condition_2}) earlier
than $f_1$, the flat direction $\phi$ starts its oscillation earlier.
Therefore, the Hubble parameter $\Hoscth$ is given by
\footnote{ We assume here hierarchical couplings and neglect effects
of the summation. }
\begin{eqnarray}
 \Hoscth &=& 
  \max_i\left[
	 {\rm min}
	 \left\{ \frac{1}{f_i^4}\frac{M_* T_R^2}{M^2}\,\,,\,\,
	  \left(c_i^2 f_i^4
	   M_* T_R^2 \right)^{1/3}\right\}
	 \right].
   \label{eq_Hosc_many}
\end{eqnarray}
Thus, $\Hoscth$ becomes larger (i.e., the oscillation begins
earlier) as we take the reheating temperature $T_R$ higher. One can see
from Eq.~(\ref{eq_Hosc_many}) that it is nontrivial which coupling among
$f_i$ makes the $\phi$'s oscillation earliest, i.e., it is not always
the coupling with strongest or weakest strength.  If one takes the
reheating temperature higher, a larger coupling makes the oscillation
earlier and determines the time $\Hoscth$.

As the universe becomes cooler, the thermal mass for the $\phi$ becomes
eventually smaller than the original mass $m$.  Then, the evolution of
$\phi$ is described by the equation of motion
\begin{eqnarray}
 \ddphi + 3 H \dphi + m^2 \phi = 0.
  \label{eq_motion_by_m}
\end{eqnarray}
The flat direction oscillates around the origin with a time scale
$\simeq m^{-1}$, and the amplitude of the oscillation is damped as
$|\phi| \sim t^{-1}\sim H$ as the universe evolves.

We have, so far, assumed that the thermal mass term dominates the
potential when the oscillation of $\phi$ starts.  However, the
expression of $\Hoscth$ in Eq.~(\ref{eq_Hosc_many}) could be smaller
than the original mass $m$, depending on the parameters $M$ and $T_R$,
which conflicts with the above assumption.  If the thermal mass does not
exceed the Hubble mass $H$ until $H \simeq m$, the original mass term
$m^2 |\phi|^2$ dominates the potential for $H\lsim m$.  Then, the
thermal effects induced by $\psi_k$ in the dilute plasma do not affect
the evolution of $\phi$.  In this case the $\phi$ field
begins its oscillation at $H \simeq m$.

In summary, the oscillation of $\phi$ starts when the Hubble parameter
of the universe becomes $H \simeq H_{\rm osc}$, where
\begin{enumerate}
 \item $H_{\rm osc} = \Hoscth$,  if the thermal mass term dominates the
       potential before $H\simeq m$,
 \item $H_{\rm osc} = m$,  if otherwise.
\end{enumerate}

\begin{figure}[t]
 \centerline{\psfig{figure=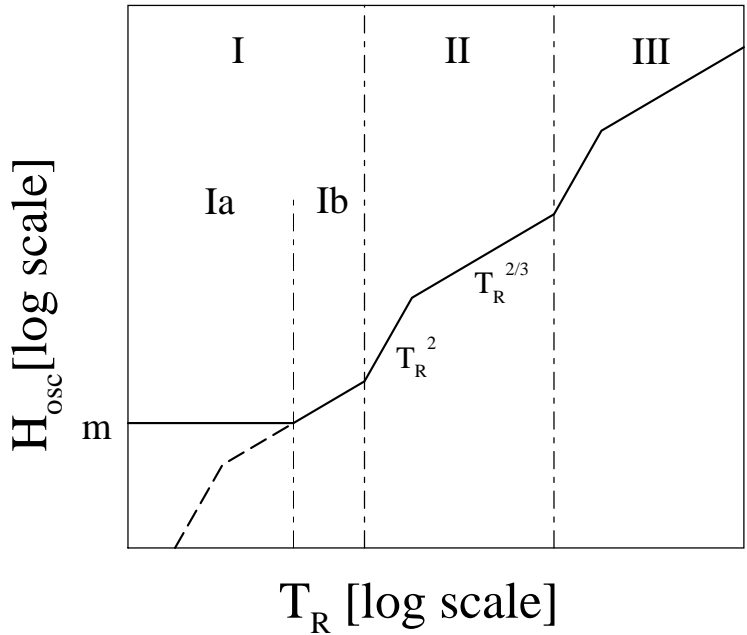,height=9.5cm}}
 \caption{ A schematic behavior of the Hubble parameter $H_{\rm
 osc}$ when $\phi$ starts its oscillation.  The oscillation of $\phi$
 starts at $H_{\rm osc} = m$ in region Ia, while the oscillation
 starts at an earlier time ($H_{\rm osc} = \Hoscth$) in regions Ib -
 III.  } 
 \label{Fig_Hosc}
\end{figure}

We show a schematic behavior of $H_{\rm osc}$ in Fig.~\ref{Fig_Hosc}
for illustration.
%
%
One can see from Eq.~(\ref{eq_Hosc_many}) that the Hubble parameter
$H_{\rm osc}$ depends on the reheating temperature non-trivially.  We
consider three couplings $f_{\rm I}$, $f_{\rm II}$ and $f_{\rm III}$
assuming a hierarchy $f_{\rm I} \ll f_{\rm II} \ll f_{\rm III}$.  The
region of $T_R$ can be divided into three regions I, II and III as in
Fig.~\ref{Fig_Hosc}, where $H_{\rm osc}^{\rm th}$ is determined by
$f_{\rm I}$, $f_{\rm II}$ and $f_{\rm III}$, respectively.  [See the
discussion below Eq.~(\ref{eq_Hosc_many}).]  The dependence of $\Hoscth$
on $T_R$ can be understood from Eq.~(\ref{eq_Hosc_1}) or
(\ref{eq_Hosc_many}).  For example, if $T_R$ lies in the region II,
$H_{\rm osc}$ behaves as
\begin{eqnarray}
 H_{\rm osc}
  \propto
  \left\{
   \begin{array}{ll}
    T_R^2     &  \mbox{for}~ T_R < {T_C}_{\rm II} \\
    T_R^{2/3} &  \mbox{for}~ T_R > {T_C}_{\rm II} \\
    \end{array}
  \right.,
\end{eqnarray}
where the critical temperature ${T_C}_{\rm II}$ is
\begin{eqnarray}
 \label{eq_TCII}
 {T_C}_{\rm II} 
  =
  f_{\rm II}^4 \left( c_{\rm II} M^3 M_\ast^{-1} \right)^{1/2}.
\end{eqnarray}
We can also see the same behavior in the regions I and III.
Furthermore, when the reheating temperature $T_R$ is low enough (i.e.,
in the region Ia), the thermal effects become negligible and the
oscillation of $\phi$ begins at $H_{\rm osc} = m$.  We see from
Fig.~\ref{Fig_Hosc} that the oscillation time $H_{\rm osc}$, which is
crucial for the estimation of the lepton asymmetry (see the next
section), depends on the reheating temperature and also the spectrum of
the couplings $f_k$ in a very complicated way.

%
\section{Lepton asymmetry}
\label{sec_lepton}
%
%
Now we are at the point to calculate the lepton asymmetry produced by the
$\phi$ field.  Since the $\phi$ field carries the lepton charge, its
number density is related to the lepton number density $n_L$ as
\begin{eqnarray}
 n_L = \frac{1}{2} i \left(
		     \dphi^* \phi - \phi^* \dphi
		     \right).
\end{eqnarray}
From Eqs.~(\ref{eq_potential}) and (\ref{eq_ofmotion}), the evolution of
$n_L$ is described by the equation,
\begin{eqnarray}
 \dot{n}_L + 3 H n_L =
  \frac{m}{2M}{\rm Im}(a \phi^4)
  + 
  \frac{H}{2M}{\rm Im}(a_H \phi^4).
  \label{eq_ofmotion-of-nL}
\end{eqnarray}
Therefore, non-trivial motion of the $\phi$ field generates the
asymmetry in lepton number~\cite{6-Affleck-Dine}.

First we consider the case $H_{\rm osc} = \Hoscth > m$.  Suppose that
the original $A$-term would vanish, i.e., $A = ma = 0$.  If this is the
case, the flat direction $\phi$ is always trapped in one of the valleys
induced by the Hubble $A$-term during the both periods $t<\toscth$ and
$t>\toscth$, and the direction of the valley does not change with
time. Therefore, there is no force which causes the motion of $\phi$
along the phase direction and no lepton-number asymmetry is produced.
However, we have the original $A$-term, and the phase of $\phi$ is
kicked by the relative phase difference between $a$ and $a_H$ in
Eq.~(\ref{eq_potential}).  The phase of $\phi$ changes during its
rolling towards the origin because the Hubble parameter $H$ decreases,
and the direction of the true valleys changes with time.
Therefore, the original $A$-term [the first term in
Eq.~(\ref{eq_ofmotion-of-nL})] plays a role of the source of the lepton
asymmetry.

One might wonder if the Hubble $A$-term [the second term in
Eq.~(\ref{eq_ofmotion-of-nL})] gives a larger contribution to the lepton
asymmetry since $H \gg m$ in the present situation.  However, since the
$\phi$ almost traces one of the valleys determined mainly by the Hubble
$A$-term, ${\rm Im}(a_H\phi^4)$ in Eq.~(\ref{eq_ofmotion-of-nL}) is
highly suppressed compared with ${\rm Im}(a\phi^4)$.  In fact, we have
found numerically that the contribution of the second term in
Eq.~(\ref{eq_ofmotion-of-nL}) is always comparable or less than that
from the original $A$-term.  Thus, we neglect the lepton asymmetry
produced from the second term in Eq.~(\ref{eq_ofmotion-of-nL}) in our
analytic calculation, for simplicity. By integrating
Eq.~(\ref{eq_ofmotion-of-nL}), we obtain the resultant lepton number at
the time $t$ as
\begin{eqnarray}
 \left[ R^3 n_L \right] (t)
  &\simeq&
  \int^t dt' R^3 \frac{m}{2M} {\rm Im}\left(a\phi^4 \right)
  \\
  &=&
  \int^{\toscth} dt' 
   R^3 \frac{m}{2M} {\rm Im}\left(a\phi^4 \right)
  +
  \int^t_{\toscth} dt' 
   R^3 \frac{m}{2M} {\rm Im}\left(a\phi^4 \right)
  ,
  \label{eq_lepton-integral}
\end{eqnarray}
where $R$ denotes the scale factor of the expanding universe, which
scales as $R\propto H^{-2/3} \propto t^{2/3}$ in the universe dominated
by the oscillating inflaton $\chi$.  Notice that the integrand of the
first term in Eq.~(\ref{eq_lepton-integral}) is almost constant, since
$|\phi| \simeq \sqrt{MH}$ and $R\propto H^{-2/3}$.  On the other hand, the
second term gives only small contribution to the total lepton asymmetry
because of the following two reasons; (i) ${\rm Im}(a\phi^4)$ changes
its sign rapidly due to the oscillation of $\phi$, and (ii) even if the
sign would remain unchanged, the amplitude of $\phi^4$ is damped as
$|\phi|^4 \sim t^{-7/2}$ (see Sec.~\ref{sec_evolution}).

Therefore, the production of lepton asymmetry is suppressed for $t
\gg \toscth$, and the resultant lepton asymmetry at $t \simeq \toscth$ is
given approximately by
\begin{eqnarray}
 n_L &=& \frac{m}{2M}{\rm Im}
     \left[ a \phi^4(\toscth)\right] 
     \toscth
 \nonumber \\
     &=&
     \frac{1}{3}m M \Hoscth \delta_{\rm eff}~,
  \label{eq_nl-tosc}
\end{eqnarray}
where $\delta_{\rm eff} \equiv \sin ( 4\arg \phi (\toscth) + \arg a)$
represents an effective CP violating phase and we assume $\delta_{\rm
eff} \simeq {\cal O}(1)$.  Notice that we have used the fact $|\phi
(t_{\rm osc}^{\rm th})| \simeq \sqrt{M \Hoscth}$.

The lepton-to-entropy ratio when the reheating process of inflation
completes ($T=T_R$) is estimated as
\begin{eqnarray}
 \frac{n_L}{s}
  =
  \frac{MT_{R}}{12 M_{\ast}^2}
  \left(\frac{m}{\Hoscth}\right)
  \delta_{\rm eff}.
  \label{eq_nl-s-1}
\end{eqnarray}
This ratio takes a constant value as long as an extra entropy production
does not take place at a later epoch, since both $n_L$ and $s$ decrease at
the same rate $R^{-3}$ as the universe expands.
In deriving Eq.~(\ref{eq_nl-s-1}) we have assumed that
the $\phi$ oscillation starts before the reheating process of inflation
completes.

Next we turn to the case without thermal effects, i.e., $H_{\rm osc}
= m$.  Detailed analysis for this case is found in
Refs.~\cite{9-Dine-Rand-Thom,15-Moro-Mura}.  Even in this case, the
production of the lepton asymmetry becomes ineffective after the
$\phi$ starts the oscillation (for $H \ll H_{\rm osc} = m$).  Thus,
the lepton asymmetry can be estimated by replacing $\Hoscth$ with $m$ in
Eq.~(\ref{eq_nl-s-1}) as
\begin{eqnarray}
 \frac{n_L}{s}
  =
  \frac{MT_{R}}{12 M_{\ast}^2}
  \delta_{\rm eff}.
  \label{eq_nl-s-2}
\end{eqnarray}

The produced lepton asymmetry is partially converted~\cite{1-Fuku-Yana}
into the baryon asymmetry through the ``sphaleron''
effects~\cite{17-sphaleron}, since it is produced before the electroweak
phase transition at $T \simeq 10^2 \GeV$. Then, the present baryon asymmetry
is given by~\cite{18-L-to-B}
\footnote{ In the present analysis we neglect the sign of the produced lepton
asymmetry.  To be exact, the lepton asymmetry should be negative, since
the present baryon asymmetry is given by~\cite{18-L-to-B}
\begin{eqnarray}
 \frac{n_B}{s} = - 
  \frac{24 + 4 N_H}{ 66 + 13 N_H} \frac{n_L}{s},
\end{eqnarray}
rather than Eq.~(\ref{eq:NB}).  Here $N_H$ denotes the number of the
Higgs doublets.  Then, $n_B/s = - (8/23) n_L/s$ in the minimal SUSY
standard model with $N_H =2$.}
\begin{eqnarray}
 \label{eq:NB}
 \frac{n_B}{s} = \frac{8}{23}\frac{n_L}{s}.
\end{eqnarray}

Now we have formulae of the lepton asymmetry Eqs.~(\ref{eq_nl-s-1}) and
(\ref{eq_nl-s-2}).  Notice that an additional suppression factor
$m/\Hoscth$ ($\Hoscth > m$) appears in Eq.~(\ref{eq_nl-s-1}), compared
with Eq.~(\ref{eq_nl-s-2}).  This suppression clearly represents the
effect of the early oscillation due to the thermal effects of the 
dilute plasma
pointed out in Ref.~\cite{9-Dine-Rand-Thom,10-Alla-Camp-Elli}.  It
should be also noted that the produced lepton asymmetry becomes larger
as the scale $M$ increases, i.e., as the neutrino mass $m_\nu$ becomes
smaller [See Eqs.~(\ref{eq_Hosc_many}),~(\ref{eq_nl-s-1}) and
(\ref{eq_nl-s-2})].  Therefore, the leptogenesis is most effective for
the flat direction corresponding to the lightest neutrino ${\nu}_1$, as
mentioned in Sec.~\ref{sec_FD}.

\begin{figure}[t]
 \centerline{\psfig{figure=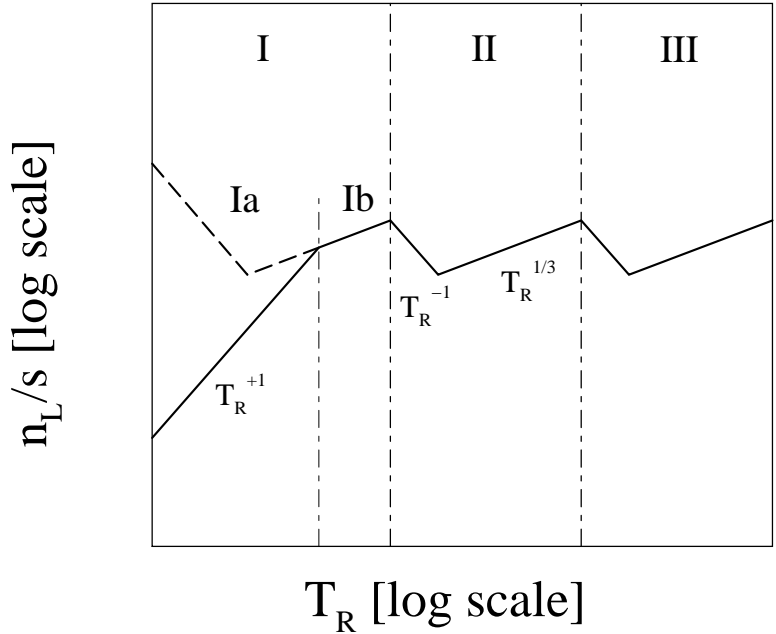,height=9.5cm}}
\caption{ A schematic behavior of the produced lepton asymmetry $n_L /
s$. Regions I, II, III correspond to those in Fig.~\ref{Fig_Hosc}.  } 
\label{Fig_TRdep_ana}
\end{figure}

The produced lepton asymmetry depends on the reheating temperature $T_R$.
We show again a schematic behavior of the lepton asymmetry in
Fig.~\ref{Fig_TRdep_ana}, where regions I, II, and III correspond to
those in Fig.~\ref{Fig_Hosc}.
%
%
If the reheating temperature $T_R$ is low enough, i.e., there is no
thermal effect ($H_{\rm osc} = m $), the lepton asymmetry $n_L / s$
is just proportional to $T_R$ as in Eq.~(\ref{eq_nl-s-2}),
\begin{eqnarray}
 \frac{n_L}{s} &\propto& T_R^{+1}.
\end{eqnarray}
However, it is not the case when the reheating temperature becomes
higher (i.e., $H_{\rm osc} = \Hoscth$), since $\Hoscth$ depends on $T_R$
(see Fig.~\ref{Fig_Hosc}).  Depending on the condition which determines
$H_{\rm osc}^{\rm th}$, the lepton
asymmetry behaves as
\begin{eqnarray}
 \label{eq:TRdep}
 \frac{n_L}{s} &\propto& T_R^{-1}
 ~{\rm or} ~ T_R^{1/3}.
\end{eqnarray}
The explicit formulae of the produced lepton asymmetry are easily
derived from Eqs.~(\ref{eq_Hosc_many}) and (\ref{eq_nl-s-1}).  For
example, if the reheating temperature lies in the region II, the
resultant lepton asymmetry is given by
\begin{eqnarray}
 \label{eq_nLsTR-ana}
 \frac{n_L}{s}
  &=&
  \frac{\delta_{\rm eff} f_{\rm II}^4 m M^3}
  {12 M_*^3 T_R}
  \qquad {\rm for} \quad T_R < {T_C}_{\rm II},
  \nonumber \\
 \frac{n_L}{s}
  &=&
  \frac{\delta_{\rm eff} m M T_R^{1/3}}
  {12 c_{\rm II}^{2/3} f_{\rm II}^{4/3} M_*^{7/3}}
  \qquad {\rm for} \quad T_R > {T_C}_{\rm II},
\end{eqnarray}
where ${T_C}_{\rm II}$ is given by Eq.~(\ref{eq_TCII}).

\begin{figure}[t]
 \centerline{\psfig{figure=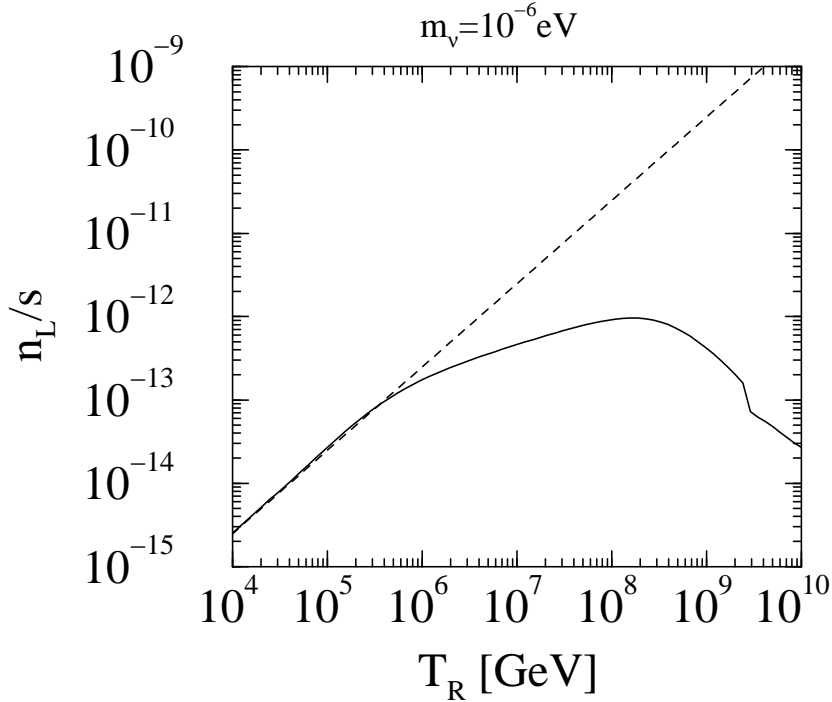,height=9.5cm}}
 \caption{ The produced lepton asymmetry $n_L/s$ for ${m_\nu}_1 =
 10^{-6}$ eV.  The solid line represents the lepton asymmetry including
 the thermal effects.  The dotted line represents the lepton asymmetry
 when one neglects the thermal effects (i.e., $f_k =0$).  We take $m =
 1$ TeV, $\tan \beta = 3$ and $\arg (a/a_H) = \pi/3$.  }
 \label{Fig_TRdep}
\end{figure}

Now, we discuss how much lepton asymmetry is generated in the SUSY
standard model.  We have performed a numerical calculation to follow the
evolution of the flat direction $\phi$ for given $M$ and $T_R$,
including all the terms in the potential in Eq.~(\ref{eq_potential}).
As for the thermal mass term, we have included all the couplings listed
in Table.\ref{table1}.  In Fig.~\ref{Fig_TRdep} we show the produced
lepton asymmetry for the lightest neutrino mass ${m_\nu}_1 = 10^{-6}$
eV.  Here, we take $m = 1$ TeV, $\tan \beta = 3$ and $\arg (a/a_H) =
\pi/3$ (we take this parameter set, hereafter).
%
%
As shown by the dotted line in Fig.~\ref{Fig_TRdep}, the produced lepton
asymmetry is proportional to the reheating temperature $T_R$, if one
neglects the thermal effects (i.e., takes all the coupling $f_k=0$). 
In this case,
we can derive from Eq.~(\ref{eq_nl-s-2}) the formula for the
lepton asymmetry
\begin{eqnarray}
 \label{eq:LAnt}
 \frac{ n_L }{ s }
 =
  0.4 \times 10^{-10} ~\delta_{\rm eff}
  \left( \frac{ T_R }{ 10^8 \GeV } \right)
  \left( \frac{ 10^{-6} \eV }{ {m_\nu}_1 } \right)~,
\end{eqnarray}
which is in a very good agreement with the result of the numerical
calculation in Fig.~\ref{Fig_TRdep}.  The above expression
Eq.~(\ref{eq:LAnt}) is also consistent with that obtained in the
previous works~\cite{9-Dine-Rand-Thom,15-Moro-Mura}.  Notice that the
produced lepton asymmetry is proportional to the reheating temperature
and inversely to the lightest neutrino mass.  Therefore, in order to
explain the baryon asymmetry in the present universe a high reheating
temperature of $T_R \simeq 10^8$ GeV is required for ${m_\nu}_1 \simeq
10^{-6}$ eV.  However, for such a high reheating temperature the effects
of the thermal mass on the dynamics of $\phi$ is significant, and hence
the $\phi$'s oscillation in fact begins at $H_{\rm osc} = \Hoscth$
rather than $H_{\rm osc} = m$. Thus, the formula Eq.~(\ref{eq:LAnt}) can
not be applied to the case of ${m_\nu}_1\simeq 10^{-6}$ eV for the high
reheating temperature.

The solid line in Fig.~\ref{Fig_TRdep} represents the lepton asymmetry
with the thermal effects.  Even in this case, the produced lepton
asymmetry is linear in $T_R$ for a low reheating temperature region of
$T_R \lesssim 4 \times 10^5$ GeV as in Eq.~(\ref{eq:LAnt}). However, we
can see that the thermal effects become significant and the lepton
asymmetry is suppressed for $T_R\gsim 10^5$--$10^6$ GeV, which is consistent with
the argument in Ref.~\cite{9-Dine-Rand-Thom}.  This is because the
thermal mass term makes the $\phi$'s oscillation earlier ($H_{\rm osc} =
\Hoscth \gg m$), which reduces the lepton asymmetry.  In fact, the
behavior of the produced lepton asymmetry shown in Fig.~\ref{Fig_TRdep}
can be understood by our analytic formula Eq.~(\ref{eq_nl-s-1}) and
Fig.~\ref{Fig_TRdep_ana}.

\begin{figure}[t]
 \centerline{\psfig{figure=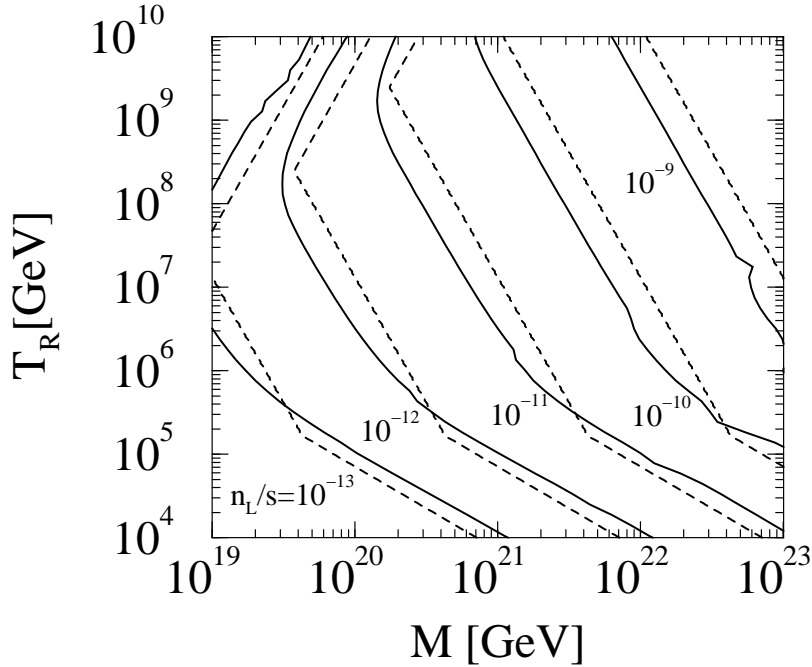,height=9.5cm}}
 \caption{ The contour plot of the lepton asymmetry $n_L/s$ in the
 $M$-$T_R$ plane.  The solid lines show the contour lines of $n_L/s$
 obtained by numerical calculation.  The dashed lines show the contour
 lines of $n_L/s$ obtained from analytic formulae.  Corresponding values of
 $n_L/s$ are also represented.  We take $m = 1$ TeV, $\tan \beta = 3$ and
 $\arg (a/a_H) = \pi/3$.  } 
 \label{Fig_contour}
\end{figure}

Fig.~\ref{Fig_contour} shows the contour plot of the 
produced lepton asymmetry $n_L / s$ in the $M$-$T_R$ plane.
%
%
We show both results by the numerical calculation and also from the
analytic formulae Eqs.~(\ref{eq_nl-s-1}) and (\ref{eq_nl-s-2}).  We
confirm that the analytic estimation discussed above reproduces very
well the result obtained by the numerical calculation.

In the lower reheating temperature region where $H_{\rm osc} = m$,
the lepton asymmetry $n_L / s$ is proportional to $M T_R$ [see
Eq.~(\ref{eq_nl-s-2})].  On the other hand, in the region where $H_{\rm
osc} = \Hoscth \gg m$, $n_L / s$ is proportional to $M^3 T_R^{-1}$ or $M
T_R^{1/3}$ [see Eqs.~(\ref{eq_Hosc_many}), (\ref{eq_nl-s-1}) and
(\ref{eq_nLsTR-ana})], depending on the condition which determines the
cosmic time $t_{\rm osc}$ when the early oscillation of $\phi$ starts.

In most of the region in Fig.~\ref{Fig_contour} the lepton
asymmetry behaves as $n_L / s \propto M T_R^{1/3}$
and the thermal effects are controlled by only the Yukawa coupling 
constant for the up quark,
which is roughly of order $10^{-5}$. Only in the parameter region $M
\lsim 10^{21}$ GeV and $T_R \gsim 10^8$ GeV, larger couplings (say, the
Yukawa coupling for the charm quark) become effective.  Note that the
result shown in Fig.~\ref{Fig_contour} is almost independent of
$\tan\beta$ as long as $\sin\beta \simeq 1$ since the Yukawa coupling
for the up quark is $y_u = m_u / \vev{H_u} = m_u / (174 {\rm GeV} \times
\sin\beta)$, where $m_u$ is the mass for the up quark.

The crucial observation is that the resultant lepton asymmetry is
suppressed by the thermal effects when the reheating temperature
is high enough.
The critical value of the reheating temperature, ${T_R}_{\rm cr}$,
is estimated analytically from Eq.~(\ref{eq_Hosc_many}) as 
\begin{eqnarray}
 \label{TRcr1}
 {T_{R}}_{\rm cr}  
  &=&
  \frac{ 1 }{ c_u f_u^2} \frac{ m^{3/2} }{ M_\ast^{1/2} } 
  =
   1 \times 10^5 \GeV
   \left( \frac{ m }{ 1 \TeV} \right)^{3/2},
\end{eqnarray}
for the region of $M \le M_{\rm cr}$, and
\begin{eqnarray}
 \label{TRcr2}
 {T_{R}}_{\rm cr}  
  &=&
  f_u^2 \frac{ M m^{1/2} }{ M_\ast^{1/2} }
   =
   4 \times 10^5 \GeV
   \left( \frac{ M }{ 10^{23} \GeV } \right)
   \left( \frac{ m }{ 1 \TeV} \right)^{1/2},
\end{eqnarray}
for the region of $M \ge M_{\rm cr}$,
where $M_{\rm cr}$ is given by
\begin{eqnarray}
 M_{\rm cr}
  \equiv
  \frac{ 1 }{ c_u f_u^4 } m
  =
  4 \times 10^{22} \GeV 
  \left( \frac{ m }{ 1 \TeV } \right).
  \label{Mcr}
\end{eqnarray}
This critical value for the scale $M$ corresponds to the lightest
neutrino mass of ${m_\nu}_1 = 0.9 \times 10^{-9}$ eV for $m =$ 1 TeV.
Here we have used the fact that only the up-quark Yukawa coupling is
effective for the thermal effects.  Therefore, we find that the thermal
effects from the dilute plasma is suppressed in the region $T_R \lesssim
10^5$--$10^6$ GeV for ${m_\nu}_1 \gtrsim 10^{-9}$ eV. This is consistent
with the result obtained in Ref.~\cite{9-Dine-Rand-Thom}.  However, for
the lighter neutrino mass region ${m_\nu}_1 \lesssim 10^{-9}$ eV, this
upper bound on the reheating temperature in order to neglect the thermal
effects increases as $M$ (i.e., inversely proportional to the lightest
neutrino mass ${m_\nu}_1$).  In fact, it helps us a lot with the
discussion in the next section.  We also confirm the result by a
numerical calculation.

In this section, we show that the thermal effects on the AD leptogenesis
is very significant.  In fact, when we require the reheating temperature
should be $T_R \lesssim 10^{8}$ GeV to avoid the gravitino
problem~\cite{7-grav-prob}, we need very a large value of $M$, say, $M
\gsim 3 \times 10^{21}$ GeV, in order to obtain the desired lepton
asymmetry $n_L / s \simeq 10^{-10}$--$10^{-9}$.  This large value $M$
corresponds to an extremely small neutrino mass ${m_\nu}_1 \lsim
10^{-8}$ eV.

%
\section{Large effective masses for right-handed neutrinos}
\label{sec:PQ}
%
%
Let us discuss the origin of the operators 
in Eq.~(\ref{eq_super}) in the
presence of the heavy right-handed Majorana neutrinos $N_I$ ($I=1,2,3$).
The relevant superpotential is
\begin{eqnarray}
 \label{eq:SP}
 W = {h_{\nu}}_{I,i}
  N_I L_i H_u + \frac{ 1 }{2} {M_{R}}_I N_I N_I~,
\end{eqnarray}
where we have used a base where the mass matrix for the heavy
neutrinos $N_I$ is diagonal.  Then, through the see-saw mechanism the
heavy Majorana neutrinos $N_I$ induce the effective operators
in Eq.~(\ref{eq_super})
at low energies,
\begin{eqnarray}
 W = 
  \frac{ 1 }{ 2 } 
  \frac{ {h_\nu}^T_{i,I} {h_\nu}_{I,j} }{ {M_R}_I} 
  L_i H_u L_j H_u~.
\end{eqnarray}
These operators gives light neutrino masses as
\begin{eqnarray}
 (m_\nu)_{i,j} 
  = v_u^2   \frac{ {h_\nu}^T_{i,I} {h_\nu}_{I,j} }{ {M_R}_I} ~.
\end{eqnarray}

It is quite natural to consider that the Majorana masses ${M_R}_I$ for
$N_I$ are given by a vacuum expectation value (vev) of some field $X$,
which is a singlet under the standard-model gauge groups, with the
superpotential
\begin{eqnarray}
 \label{eq:MR}
 W = \frac{1}{2}{h_N}_I X N_I N_I~.
\end{eqnarray}
If it is the case, masses of the heavy Majorana neutrinos $N_I$ in the early
universe may be different from those in the present universe. Namely,
the extremely small neutrino mass discussed in the previous section
(${m_\nu}_1 \lsim 10^{-8}$ eV) may not be the same as the observed mass.
In particular, when the field $X$ is responsible to the Peccei-Quinn
symmetry breaking~\cite{11-Peccei-Quinn}, we have necessarily a flat
direction containing the $X$ field.  Thus, the $X$ field may have a
large value as $X \simeq M_{\ast}$ during the inflation whereas it has a
vev of $\vev{X} \simeq F_a$ in the true vacuum. Here, the Peccei-Quinn
breaking scale $F_a$ is constrained by laboratory experiments,
astrophysics, and cosmology as $F_a \simeq 10^{10}$--$10^{12}$
GeV~\cite{19-axion-Kim}.
\footnote{
The upper bound on $F_a$ is relaxed as $F_a \lesssim 10^{16}$ GeV,
when there is an extra entropy production after the QCD phase
transition~\cite{20-Fa}.
}
In this case, neutrino masses in the early universe are much smaller
than those observed today (they are suppressed by the factor $X /F_a
\simeq M_{\ast}/F_a \simeq 10^6$--$10^8$).  Therefore, the extremely
small neutrino mass required for a successful leptogenesis may be
naturally explained in this scenario.

To demonstrate our point, we consider the following superpotential for
the Peccei-Quinn symmetry breaking sector,
\begin{eqnarray}
 \label{eq:SP2}
 W = \lambda Y ( X \overline{X} - F_a^2 )~,
\end{eqnarray}
where $\lambda$ is a coupling constant, and $Y$, $X$, and $\overline{X}$
are supermultiples which are singlets under the standard-model gauge
groups and have 0, +1, and $-1$ Peccei-Quinn charges, respectively.  From
Eq.~(\ref{eq:SP2}) we see that there exists a flat direction $X
\overline{X} = F_a^2$. We parameterize this direction by the
scalar field $\sigma$ called ``saxion''.  This saxion $\sigma$ receives
a soft SUSY breaking mass of $m_\sigma \simeq m_{3/2}$
($m_{3/2}$ is the gravitino mass), and the $X$ and
$\overline X$ have vevs of the order of $F_a$ if the
soft SUSY breaking masses for $X$ and $\overline{X}$ are almost the
same. (${\cal L} = m_X^2 |X|^2 + m_{\overline X} |\overline{X}|^2$ with
$m_X \simeq m_{\overline X} \simeq m_{3/2}$).

Along this flat direction the $X$ field may have a large value during
the inflation, if the additional SUSY breaking effects induce a negative
mass squared for $X$. In this case the supergravity effects prevent the
$X$ field from running over $M_\ast$ and give an initial value $X \simeq
M_{\ast}$~\cite{21-Kasuya-Kawasaki-Yanagida}.  We here assume $X\simeq
M_*$ and hence $\overline{X} \simeq F_a^2/M_{\ast}$ at the end of
inflation.
\footnote{ In this case, the isocurvature fluctuation
associated with the axion is very
suppressed~\cite{21-Kasuya-Kawasaki-Yanagida}.  }

We turn to calculate the produced lepton asymmetry in the presence of
the $X$ field.  We see from Eq.~(\ref{eq:MR}) that the masses for $N_I$
are
\begin{eqnarray}
 {M_R}_I = {h_N}_I X~.
\end{eqnarray}
Then, we find
\begin{eqnarray}
 (m_\nu)_{i,j} 
  = \frac{ v_u^2 }{ X }   
    {h_\nu}^T_{i,I} {h_N}_I^{-1} {h_\nu}_{I,j}  ~.
\end{eqnarray}
The vev of $X$ in the true vacuum, $\langle X \rangle \simeq F_a$, is
supposed to give the neutrino masses observed today.  As discussed in
the previous sections, the lightest neutrino $\nu_1$ ($ {m_\nu}_1 \ll
{m_\nu}_2 \ll {m_\nu}_3$) is relevant for the AD leptogenesis.

As for the neutrino masses we adopt the recent data from 
the Superkamoikande experiments.  
The atmospheric neutrino oscillation is well explained by
the $\nu_\mu$--$\nu_\tau$ oscillation with ${m_\nu}_3 \simeq 5 \times
10^{-2}$ eV~\cite{2-Kamioka-atm}. On the other hand, the large MSW
solution to the solar neutrino problem is favored by the recent Superkamiokande
data~\cite{12-Suzuki}, which indicates that ${m_\nu}_2 \simeq 5 \times
10^{-3}$ eV.  However, it is very difficult to determine the lightest
neutrino mass ${m_\nu}_1$.  Here, we simply take ${m_\nu}_1 = 1 \times
10^{-4}$ eV for a representative value.

Then, the scale $M$ in the previous sections is estimated as
\begin{eqnarray}
 \label{eq:M1}
  M = 3 \times 10^{17}  \GeV
  \left( \frac{ 10^{-4} \eV}{ {m_\nu}_1 } \right)
  \left( \frac{ \langle X \rangle }{ F_a } \right) ~.
\end{eqnarray}
Notice that as shown in Fig.~\ref{Fig_contour} such a scale $M$ for
$\vev{X}\simeq F_a$ can not produce sufficient lepton asymmetry to
explain the baryon asymmetry in the present universe.  However, the
large initial value of $X$ ($X_0 \simeq M_{\ast}$) gives
\begin{eqnarray}
 \label{eq:M1f}
  M 
  =
   7 \times 10^{25}  \GeV
   \left( \frac{ 10^{-4} \eV}{ {m_\nu}_1 } \right)
   \left( \frac{ 10^{10} \GeV }{ F_a } \right)
   \left( \frac{ X_0 }{ M_{\ast} } \right) ~,
\end{eqnarray}
which is translated into the lightest neutrino mass in the false vacuum
\begin{eqnarray}
 \label{eq:mnu1f}
  {m_\nu}_1' = 4 \times 10^{-13} \eV
   \left( \frac{ {m_\nu}_1 }{ 10^{-4} \eV }\right)
   \left( \frac{ F_a }{ 10^{10} \GeV }\right)
   \left( \frac{ M_{\ast} }{ X_0 } \right).
\end{eqnarray}
Thus, such a high scale $M$ in Eq.~(\ref{eq:M1f}) can give a
neutrino mass much smaller than that observed today.  It should be noted
that the flat direction saxion $\sigma$ stays at $X_0 \simeq M_{\ast}$
until $H \simeq m_\sigma \simeq m_{3/2}$ due to the friction of the
expansion of the universe.  
Furthermore, we find from Eqs.~(\ref{TRcr1}) and (\ref{TRcr2}) 
that for such a high scale of $M \simeq 10^{26}$ GeV 
we can estimate the lepton asymmetry without the thermal effects
even if the reheating temperature is as 
high as $T_R \simeq 10^8$ GeV.

Therefore, for the reheating temperature of $T_R \lesssim 10^8$ GeV,
the produced lepton asymmetry is estimated by using
Eq.~(\ref{eq_nl-s-2}) as
\begin{eqnarray}
 \frac{ n_L}{s} &=& 
 \frac{T_R}{12 M_*^2} 
 \left( \frac{v_u^2 X_0}{{m_\nu}_1 F_a}\right)
 \delta_{\rm eff}
 \nonumber \\
 &=&
 1 \times 10^{-4} ~\delta_{\rm eff}
 \left( \frac{ T_R }{10^8 \GeV } \right)
 \left( \frac{ 10^{-4} \eV }{ {m_\nu}_1 } \right)
 \left( \frac{ 10^{10} \GeV }{ F_a } \right)
 \left( \frac{ X_0 }{ M_{\ast} } \right)~. 
 \label{eq:PLA}
\end{eqnarray}
This shows that too large amount of the lepton asymmetry is produced
through the AD mechanism. However, this asymmetry is sufficiently
diluted, since there exists substantial entropy production at low
energies.  The saxion $\sigma$ is produced in the coherent oscillation
after the lepton asymmetry is frozen, and the oscillation energy
dominates the universe.
\footnote{
The saxions are also produced by the thermal scatterings in the
reheating period.  However, the energy density of the saxions produced
thermally is much smaller than that produced in the coherent
oscillations, and hence we neglect it.
}
Then, the late decay of the
saxion increases the entropy of the universe and dilutes the lepton
(baryon) asymmetry substantially.

The saxion begins the coherent oscillation at $H \simeq m_\sigma \simeq
m_{3/2}$ with the initial amplitude $X_0 \simeq M_*$. The oscillation
energy at that time is given by $\rho_\sigma \simeq m_\sigma^2 X_0^2
/2$.  Note that the saxion oscillation starts before the reheating
process of the inflation takes place.
\footnote{ It is the case for $T_R \lesssim 2 \times 10^{10} \GeV ( m_\sigma
/ 1 \TeV)^{1/2}$.  }
Then, when $T = T_R$, the ratio of $\rho_\sigma$ to the entropy density
of the universe is estimated as
\begin{eqnarray}
 \label{ABs1}
 \frac{ \rho_\sigma }{ s }
  =
  \frac{ 1 }{ 8 }
  T_R
  \left( \frac{ X_0 }{ M_{\ast} }\right)^2~.
\end{eqnarray}
Notice that $\rho_\sigma$ decreases at the rate $R^{-3}$ as the universe
expands, while the radiation energy density decreases as $R^{-4}$.
Therefore, the oscillation energy of the saxion dominates the energy of
the universe soon after the reheating process completes.  This energy is
transferred into thermal bath by the saxion decay.  The saxion decays
into two gluons with the partial decay rate
\begin{eqnarray}
 \Gamma (\sigma \rightarrow 2g )
  =
  \frac{ \alpha_s^2 }{ 32 \pi^3 }
  \frac{ m_\sigma^3 }{ F_a^2 }~.
\end{eqnarray}
Through this decay the universe is reheated again, and its reheating
temperature $T_\sigma$ is estimated as
\begin{eqnarray}
 \label{Ts}
 T_\sigma =
  10  \GeV
  \left( \frac{ m_\sigma}{1 \TeV}\right)^{3/2}
  \left( \frac{ 10^{10} \GeV}{ F_a } \right).
\end{eqnarray}
The saxion decay takes place far before the beginning
of the big-bang neucleosynthesis (BBN), and hence is cosmologically
harmless.  It should be noted that the saxion might decay dominately
into two axions. If it is the case, the extra energy of the axion at the
BBN epoch raises the Hubble expansion of the universe, which leads to
overproduction of $^4$He.  To avoid this difficulty the branching ratio
of the saxion decay into two axion should be smaller than about 0.1.
Here, we simply assume that the dominant decay process of saxion is
$\sigma \rightarrow 2 g$.
\footnote{ This is realized when $m_X^2 \simeq m_{\overline{X}}^2$ .}
Then the saxion decay increases the entropy
of the universe by the rate
\begin{eqnarray}
 \label{eq:EP}
 \Delta 
  &=& \frac{ T_R }{ 6 T_\sigma}
  \left( \frac{ X_0 }{ M_{\ast}} \right)^2
  \nonumber \\
  &=&
   2 \times 10^6 
   \left( \frac{ T_R }{10^8 \GeV } \right)
   \left( \frac{ 1 \TeV }{ m_\sigma } \right)^{3/2}
   \left( \frac{ F_a }{ 10^{10} \GeV } \right)
  \left( \frac{ X_0 }{ M_{\ast}} \right)^2 ~.
\end{eqnarray}
Because of this entropy production by the saxion decay, the primordial
lepton asymmetry shown in Eq.~(\ref{eq:PLA}) is also diluted by the rate
$\Delta$.  Then the present baryon asymmetry is given by
\begin{eqnarray}
 \frac{ n_B }{s}
  &=&
  4 \times 10^{-4}~ \delta_{\rm eff}
  \frac{ v_u^2 m_\sigma^{3/2} }
  { {m_\nu}_1 F_a^2 M_{\ast}^{1/2} }
  \left( \frac{ M_{\ast} }{ X_0 } \right)
  \nonumber \\
 &=&
  0.2 \times 10^{-10} ~ \delta_{\rm eff}
  \left( \frac{ 10^{-4} \eV }{ {m_\nu}_1 } \right)
  \left( \frac{ 10^{10} \GeV }{ F_a } \right)^2
  \left( \frac{ m_\sigma }{ 1 \TeV } \right)^{3/2}
  \left( \frac{ M_{\ast} }{ X_0 } \right)~.
  \label{NB_PQ}
\end{eqnarray}
Notice that the present baryon asymmetry is independent on the reheating
temperature $T_R$.  We see that the desired baryon asymmetry is just
given by the lightest neutrino mass of ${m_\nu}_1 \simeq 10^{-4}$ eV for
$F_a \simeq 10^{10}$ GeV.

Before closing this section, we should comment on the
cosmological consequence of our model.  The entropy production by the
saxion in Eq.~(\ref{eq:EP}) ensures that we are free from the
cosmological gravitino problem.  The number density of gravitinos
produced at the reheating process is diluted by the rate 
$\Delta$ and hence the radiative decay of gravitino dose not 
disturb the BBN. Furthermore, we find that the model does not
suffer from the cosmological problem of the axino which is a
fermionic superpartner of the axion.  The axinos are produced in the
reheating process by the thermal scatterings and may lead to
a cosmological difficulty~\cite{22-Rajagopal-Turner-Wilczek}.  However,
in our model, the interaction of axino at the reheating epoch is
suppressed by $X_0 \simeq M_{\ast}$, not by $F_a$, and hence the
production of axino is less effective.  Furthermore, the entropy
production by the saxion decay dilutes the axino abundance 
and hence the axino
becomes completely cosmologically harmless.

%
\section{Conclusions and discussion}
%
In this paper we have performed a detailed analysis on the Affleck-Dine
leptogenesis taking into account the thermal effects from the dilute
plasma.  We have first shown that the thermal effects change drastically
the dynamics of the flat direction $\phi$ and suppress the lepton-number
asymmetry produced by the flat direction $\phi$.  In order to escape
from the thermal effects, a relatively low reheating temperature of $T_R
\lesssim 10^5$--$10^6$ GeV is found to be required when the lightest
neutrino mass is ${m_\nu}_1 \gtrsim 10^{-9}$ eV.  This is consistent
with the result obtained in Ref.~\cite{9-Dine-Rand-Thom}.  On the other
hand, we have found that this upper bound on $T_R$ is relaxed as $T_R
\lesssim 10^5$--$10^6$ GeV $\times ( {m_\nu}_1 / 10^{-9} \eV)^{-1}$ for
the lighter neutrino mass region of ${m_\nu}_1 \lesssim 10^{-9}$ eV.  We
have also estimated the resultant lepton asymmetry for the higher
reheating temperature region where the thermal effects are important by
both analytical and numerical calculations.  We have found that an
ultralight neutrino with a mass such as ${m_\nu}_1\simeq 10^{-8}$ eV is
required to produce enough lepton asymmetry to account for the baryon
asymmetry in the present universe.  Here, we have assumed that the
reheating temperature should be lower than about $10^8$ GeV to avoid the
cosmological problem of the gravitino of mass $m_{3/2}\simeq 100$ GeV--1
TeV.  Such an ultralight neutrino seems to be very unlikely the case,
since the recent Superkamiokande
experiments~\cite{2-Kamioka-atm,12-Suzuki} suggest the masses of heavier
two neutrinos $\nu_2$ and $\nu_3$ to be in a range of
$10^{-1}$--$10^{-3}$ eV.

However, in the second part of this paper, we have pointed out that the
above neutrino mass ${m_\nu}_1\simeq 10^{-8}$ eV is not necessarily
the mass to be observed today, if the heavy Majorana masses of the
right-handed neutrinos are dynamical valuables in the early universe. We
construct a model based on the Peccei-Quinn symmetry to demonstrate our
point. We find in this model that the neutrino mass ${m_\nu}_1$ in the
true vacuum can be as large as ${m_\nu}_1\simeq 10^{-5}$--$10^{-4}$ eV
to obtain $n_B / s\simeq 10^{-10}$--$10^{-11}$ for the reheating
temperature of $T_R \lsim 10^8$ GeV. Therefore, we have shown that the AD
leptogenesis works well as long as the masses for the right-handed
neutrinos are dynamical valuables in the early universe.

\section*{Acknowledgements}

This work was partially supported by the Japan Society for the
Promotion of Science (T.A. and K.H.) and ``Priority Area: Supersymmetry
and Unified Theory of Elementary Physics ({\#} 707)'' (T.Y.).
%


%
%
%
%
\newcommand{\Journal}[4]{{\sl #1} {\bf #2} {(#3)} {#4}}
\newcommand{\PL}{\sl Phys. Lett.}
\newcommand{\PR}{\sl Phys. Rev.}
\newcommand{\PRL}{\sl Phys. Rev. Lett.}
\newcommand{\NP}{\sl Nucl. Phys.}
\newcommand{\ZP}{\sl Z. Phys.}
\newcommand{\PTP}{\sl Prog. Theor. Phys.}
\newcommand{\NC}{\sl Nuovo Cimento}
\newcommand{\MPL}{\sl Mod. Phys. Lett.}
\newcommand{\PRep}{\sl Phys. Rep.}

\end{document}